\documentclass[prc,twocolumn,superscriptaddress,showpacs,preprintnumbers,amsmath,amssymb]{revtex4}
\usepackage[dvips]{graphicx}
\def\m@thcombine#1#2{%
  \setbox0=\hbox{$#1$}
  \setbox1=\hbox{$#2$}
  \ifdim\wd0>\wd1
    \setbox0=\hbox to\wd1{\hss\box0\hss}
  \else
    \setbox1=\hbox to\wd0{\hss\box1\hss}
  \fi
  \mathop{\vcenter{
    \offinterlineskip\box0\box1}}}
\def\lesim{\m@thcombine<\sim}
\def\gesim{\m@thcombine>\sim}
\def\vr{\mbox{\boldmath$r$}}

\begin{document}
\title{Constraining slope parameter of symmetry energy from nuclear structure}
\author{T. Inakura}
\affiliation{Department of Physics, Graduate School of Science, Chiba University, Chiba 263-8522, Japan}
\affiliation{Yukawa Institute of Theoretical Physics, Kyoto University, Kyoto 606-8502, Japan}
\affiliation{Department of Physics, Niigata University, Niigata 950-2181, Japan}
\author{H. Nakada}
\affiliation{Department of Physics, Graduate School of Science, Chiba University, Chiba 263-8522, Japan}

\begin{abstract}
Four quantities deducible from nuclear structure experiments
have been claimed to correlate to the slope parameter $L$ of the symmetry energy;
the neutron skin thickness, the cross section of low-energy dipole (LED) mode,
dipole polarizability $\alpha_D$,
and $\alpha_D S_0$ (i.e. product of $\alpha_D$ and the symmetry energy $S_0$).
By the calculations in the Hartree-Fock plus random-phase approximation
with various effective interactions,
we compare the correlations between $L$ and these four quantities.
The correlation derived from different interactions
and the correlation from a class of interactions that are identical in the symmetric matter as well as in $S_0$
are simultaneously examined.
These two types of correlations may behave differently,
as exemplified in the correlation of $\alpha_D$ to $L$.
It is found that the neutron skin thickness and $\alpha_DS_0$
correlate well to $L$, and therefore are suitable
for narrowing down the value of $L$ via experiments.
The LED emergence and upgrowth makes the $\alpha_DS_0$-$L$ correlation strong,
although these correlations are disarranged
when neutron halo appears in the ground state.

\end{abstract}

\pacs  {
 21.65.-f, 
 21.65.Mn, 
 24.30.Cz, 24.30.Gd, 
 21.60.Jz, 
 25.20.-x 
}
\maketitle

\section{Introduction}

Properties of nuclear matter is a basic subject in nuclear physics.
The equation of state (EoS) of the symmetric nuclear matter (SNM), which is characterized 
by the saturation density $\rho_0$, the saturation energy $E/A(\rho_0)$ and the 
incompressibility $K_\infty$, has been studied for a long time and its properties 
around $\rho_0$ are known rather well.
In contrast, the EoS of the pure neutron matter (PNM) has not been established,
despite its importance connected with compact astrophysical objects,
e.g. neutron stars (NSs).
Recent observation of a two-solar-mass (2$M_\odot$) NS~\cite{Demorest10} 
has imposed a constraint on the EoS,
and has given an additional momentum for resolving the PNM EoS in particular.
Based on the SNM EoS, the PNM EoS is mostly governed by the symmetry energy $S$
as a function of density $\rho$,
which is characterized by $S_0 = S(\rho=\rho_0)$ and the slope parameter, 
\begin{eqnarray}
L = 3 \rho_0 \left. \frac{\partial S(\rho)}{\partial \rho} \right|_{\rho=\rho_0} \,.
\end{eqnarray}
As $S_0$ has long been investigated and is known rather well,
the current uncertainty in the PNM EoS mainly originates
in the uncertainty in $L$.

Although pure neutron many-body systems do not exist on earth,
experiments using radioactive beams disclosed
that many nuclei have certain volumes dominated by neutrons;
i.e. neutron skins.
This may open a possibility to constrain the PNM EoS
from experiments on structure of the neutron-rich nuclei.
Objects dominated by neutrons may be formed also in the process of nuclear reactions,
which could leave a signal in observables.
Many studies narrowing the PNM EoS have been devoted to 
searching observables which strongly correlate with $L$; e.g.
nuclear mass systematics~\cite{Danielewicz09,Kortelainen10,Liu10,Dong12,Moeller12}, 
neutron skin thickness~\cite{Centelles09,Warda09,Chen10,Zenihiro10,RocaMaza11,Agrawal12}, 
fragmentation in the heavy ion collisions~\cite{Chen05,Famiano06,Shetty07,Tsang09}, 
and low-lying $E1$ mode (LED) \cite{Carbone10,Inakura13} in unstable nuclei.
Among them, we focus on quantities relevant to structure of specific nuclides,
for which model-dependence is considered to be relatively weak.

In Ref.~\cite{RocaMaza11},
the neutron skin thickness $\Delta r_{np}$ in $^{208}$Pb
has been found to correlate linearly to $L$ with a large correlation coefficient $0.98$,
by calculations using 47 effective interactions.
This suggests that accurate determination of $\Delta r_{np}$ 
serves constraining $L$.
The LED mode is considered as a relative oscillation
between the neutron skin and the remnant core.
In Ref.~\cite{Carbone10}, a linear correlation
between the LED cross section ($\sigma_\mathrm{LED}$) and $L$ has been suggested,
from calculations in the random-phase approximation (RPA) for $^{68}$Ni and $^{132}$Sn with 26 effective interaction.
By combining it with the experimental data,
$L=49-81$\,MeV has been deduced~\cite{Wieland09,Adrich05}.
However, the covariance analysis for effective interactions~\cite{Reinhard10,Paar14,RocaMaza14}
has shown that this correlation is not always strong.
Instead, the dipole polarizability $\alpha_D$ has been claimed to be better in constraining $L$
than cross section and transition strength of the LED.
If the $\alpha_D$-$L$ correlation is assumed,
the experimental data in $^{208}$Pb indicate $L=46 \pm 15$\,MeV~\cite{Tamii11}.
It has further been argued, in Ref.~\cite{RocaMaza13},
that a product of $\alpha_D$ and $S_0$ is better correlated with $L$ than $\alpha_D$ alone,
based on the droplet model with some assumptions.

The above four quantities ($\Delta r_{np}$, $\sigma_\mathrm{LED}$, $\alpha_D$ and $\alpha_D S_0$)
have been proposed in separate works,
and there have been few studies comparing them directly,
with exception of Ref.~\cite{Reinhard13}. 
Moreover, depending on the studies,
two different types of the correlations have been argued
that should be distinguished.
The $\alpha_D$-$L$ correlation has been investigated using the covariance analysis,
for which a single interaction and its variants are employed.
These variants are generated so as to have similar properties to the original interaction except $L$.
In contrast, the other correlations have been investigated using many interactions
with different origin.
It is not obvious whether these two types of correlations have the same behavior.
We also point out that nucleus-dependence has not been discussed sufficiently.
Most calculations have been implemented in $^{68}$Ni, $^{132}$Sn and $^{208}$Pb,
partly because they are spherical, neutron-rich and accessible by experiments.
Nuclear deformation possibly draws complication, indeed.
Still, there could be better candidates.
Further investigation including careful assessment of correlations is desired
in order to constrain $L$ from experimental data.

In this article we investigate the correlations of $\Delta r_{np}$, $\sigma_\mathrm{LED}$, $\alpha_D$ 
and $\alpha_D S_0$ with $L$ for a number of spherical nuclei. The paper is organized as follows. 
In Sec.~\ref{sec:Method}, we briefly explain interactions we employ and introduce an additional term 
to them, which controls the value of $L$.
Numerical results are given in Sec.~\ref{sec:result},
and we discuss the interaction- and nucleus-dependence of the correlations.
Conclusion is given in Sec.~\ref{sec.conclusion}.

\section{Method}
\label{sec:Method}


We perform the RPA calculations on top of the Hartree-Fock (HF) wave functions
in fully self-consistent manner,
by using the numerical methods of Refs.~\cite{Inakura09,Nakada09}.

In investigating interaction-dependence of the correlations between $L$ and the quantities, 
we employ a variety of effective interactions, covering a wide range of $L$.
They are three Skyrme interactions which have widely been used
(SkM$^\ast$~\cite{SkM*}, SLy4~\cite{SLy4} and SGII~\cite{SGII}), 
two latest designed ones (UNEDF0 and UNEDF1~\cite{UNEDF}),
and four Skyrme interactions (SkI2, SkI3, SkI4 and SkI5~\cite{SkIseries}) that give large $L$ values,
and two more Skyrme interactions (SkT4~\cite{SkT4} and Ska~\cite{Ska})
which are less frequently used but useful for checking robustness of the correlations.
In addition, three Gogny (D1~\cite{GognyD1}, D1S~\cite{GognyD1S} and D1M~\cite{GognyD1M}) and 
two M3Y-type interactions (M3Y-P6 and M3Y-P7~\cite{M3Y}) are adopted.
Using these effective interactions which cover $L= 18-129$\,MeV, we discuss the correlations among different interactions (CDI).
There have been a certain number of relativistic mean-field (RMF) calculations.
Most of the RMF Lagrangians adopted so far tend to give large $L$ values ($\gtrsim 100$\,MeV),
which do not seem compatible with experimental data.
Their results are similar, though not identical, to the SkI$n$ ($n=2-5$) ones.
There may be rooms to obtain RMF Lagrangians giving smaller $L$ values.
Although we have not implemented the RMF calculations,
we shall mention some of the RMF results available in literature.

\begin{figure}[tb]
\begin{center}
\includegraphics[width=0.4990\textwidth,keepaspectratio]{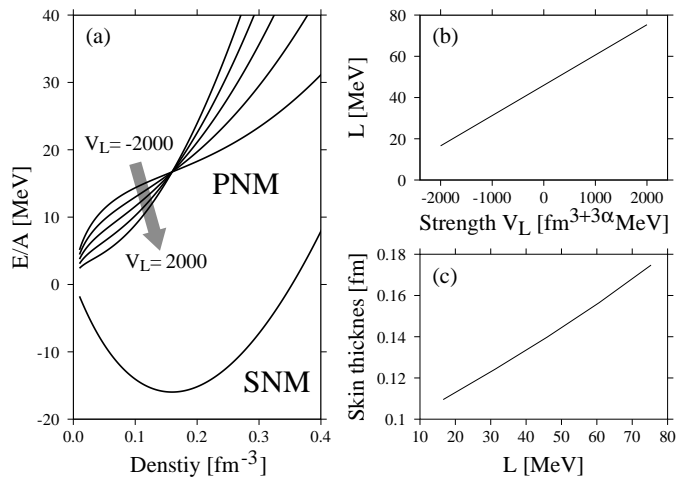}
\caption{$V_L$ dependence (Eq.~(\ref{VL})) of (a) EoS and (b) the slope 
parameter $L$, calculated with SLy4 interaction on setting 
$V_L= 0,\, \pm 1000, \,\pm 2000$\,fm$^{3+3\alpha}$MeV, and (c) relation between 
neutron skin thickness in $^{208}$Pb and $L$ shifted by adjusting $V_L$.}
\label{Ldep.208Pb}
\end{center}
\end{figure}

In the covariance analysis in Refs.~\cite{Reinhard10,Paar14,RocaMaza14},
a class of interactions that share basic properties with an original interaction were considered.
Following Ref.~\cite{Ono03},
we here introduce an additional term for the interaction,
\begin{equation}
v_{ij} \Longrightarrow v_{ij} - V_L \left[ \rho^\alpha(\vr_i) - \rho^\alpha_0 \,\right] P_\sigma \delta( \vr_i - \vr_j ) \,,
\label{VL}
\end{equation}
where $P_\sigma$ is the spin exchange operator.
This additional term does not change $S_0$ because it vanishes at $\rho=\rho_0$,
and has no effects on the SNM EoS
because $\langle P_\sigma \delta( \vr_i - \vr_j )\rangle=0$ in the SNM.
We thus obtain a class of interactions having different $L$ by varying $V_L$,
with changing neither SNM EoS nor $S_0$. 
All the non-relativistic interactions contain a density-dependent term
in which the coupling constant is proportional to a power of the density.
We keep this power $\alpha$ of each original interaction
also for the additional term in Eq.~(\ref{VL}).
The correlation given by the interactions belonging to the same class,
which are generated from a single interaction but have different $V_L$, 
will be called correlation in a single class of interactions (CSI)
in this paper.

Figure~\ref{Ldep.208Pb}(a) illustrates how $V_L$ affects the EoS, 
by taking the SLy4 interaction and its variants with $V_L= 0, \pm 1000, \pm 2000$\,fm$^{3+3\alpha}$MeV as an example.
The $L$ value is changed linearly with $V_L$, as $V_L= -2000$\,$(2000)$\,fm$^{3+3\alpha}$MeV shifts $L$ from the original value 46\,MeV to $17\,(75)$\,MeV.
The neutron skin thickness is defined by
\begin{equation}
\Delta r_{np} = \sqrt{\langle r^2 \rangle_n} - \sqrt{\langle r^2 \rangle_p},
\end{equation}
for a specific nuclide.
As is expected, $L$ correlates linearly with the neutron skin thickness
in $^{208}$Pb among this class of interactions,
as shown in Fig.~\ref{Ldep.208Pb}(c). 
The additional term changes the binding energy of $^{208}$Pb by $\sim 15$\,MeV
with $V_L= -2000\,(2000)$\,fm$^{3+3\alpha}$MeV.
We do not take this difference seriously,
since this energy shift is comparable to the difference of the binding energies
obtained from different interactions.
For instance, UNEDF0 and UNEDF1 yield 1625 and 1643\,MeV for $^{208}$Pb, respectively.

The $E1$ transition operator is expressed as
\begin{equation}
\mathcal{O}^{(E1)} = \frac{N}{A} \sum_{i \in p} r_i Y^{(1)}(\Omega_i) - \frac{Z}{A} \sum_{i \in n} r_i Y^{(1)}(\Omega_i) \,,
\end{equation}
after the center of mass correction. Here $i$ is the index of nucleons and $i \in p$ ($i\in n$)  indicates 
that the sum runs over protons (neutrons). 
The $E1$ strength is calculated as
\begin{eqnarray}
&& S^{(E1)}(\omega) = \frac{\gamma}{\pi} \sum_n \left[ \frac{1}{\left( \omega-\omega_n \right)^2 + \gamma^2} -  \frac{1}{\left( \omega+\omega_n \right)^2 + \gamma^2} \right] \nonumber \\
&& \phantom{S^{(E1)}(\omega)= \frac{\gamma}{\pi} \sum_n } \qquad \times \left| \langle \Phi_n | \mathcal{O}^{(E1)} | \Phi_0 \rangle \right|^2
\label{E1strength}
\end{eqnarray}
where $n$ is the index of the excited states and $\omega$ denotes the excitation energy.
For the smearing parameter $\gamma$, we adopt $\gamma=0.5$\,MeV,
after confirming that the results do not change much with $\gamma = 0.1-0.5$\,MeV.
The LED cross section $\sigma_\mathrm{LED}$ is given by
\begin{eqnarray}
\sigma_\mathrm{LED} = \frac{16\pi^3 e^2}{9\hbar c}\int^{\omega_\mathrm{dip}}_0 \!\!\!\! d\omega\, \omega S^{(E1)}(\omega) \,,
\label{crosssection}
\end{eqnarray}
where $\omega_\mathrm{dip}$ is the energy at which $S^{(E1)}(\omega)$ is separated 
into the LED and giant dipole resonance (GDR) regions.
Although the LED and the GDR components could mix in certain energy range~\cite{Nakada13},
we here separate them by energy for simplicity.
It is not obvious how $\omega_\mathrm{dip}$ should be defined.
We determine $\omega_\mathrm{dip}$ as follows.
If we find a distinguishable LED peak in $S^{(E1)}(\omega)$,
$\omega_\mathrm{dip}$ is defined as the energy corresponding to the minimum 
of $S^{(E1)}(\omega)$ that exists between the LED peak and the GDR.

The dipole polarizability $\alpha_D$ is calculated as
\begin{eqnarray}
\alpha_D = \frac{8\pi e^2}{9}\int^\infty_0 \!\! d\omega\, \frac{S^{(E1)}(\omega)}{\omega} \,.
\label{polarizability}
\end{eqnarray}
Owing to the energy denominator,
$\alpha_D$ is expected to be sensitive to the LED.
It should be noted that $\alpha_D$ is unambiguously defined unlike $\sigma_\mathrm{LED}$.

As a measure of correlations, it is customary to use the correlation coefficient.
For the two quantities $(x,y)$ for which we have data points $(x_k,y_k)$
($k=1,2,\cdots,N_d$), the correlation coefficient is given by
\begin{equation}
 R[x,y]=\frac{\displaystyle\sum_{k=1}^{N_d}(x_k-\bar{x})(y_k-\bar{y})}
 {\sqrt{\displaystyle\sum_{k=1}^{N_d}(x_k-\bar{x})^2}
 \sqrt{\displaystyle\sum_{k=1}^{N_d}(y_k-\bar{y})^2}}\,, \label{corrcoef}
\end{equation}
with $\bar{x}=\sum_{k=1}^{N_d} x_k/N_d$ and likewise for $\bar{y}$.
We obtain $|R[x,y]|=1$ if $x$ and $y$ are fully correlated
and $R[x,y]=0$ if $x$ and $y$ are fully uncorrelated.
In the present case $k$ corresponds to individual interaction,
covering the interactions mentioned above including the variants with varying $V_L$.
$x$ is fixed to be the slope parameter $L$,
and $y$ is taken to be $\Delta r_{np}$, $\sigma_\mathrm{LED}$, $\alpha_D$
or $\alpha_DS_0$ for a specific nuclide.

\section{Numerical Results}
\label{sec:result}

\subsection{Correlations in $^{132}$Sn}
\begin{figure}[!tb]
\begin{center}
\includegraphics[width=0.4990\textwidth,keepaspectratio]{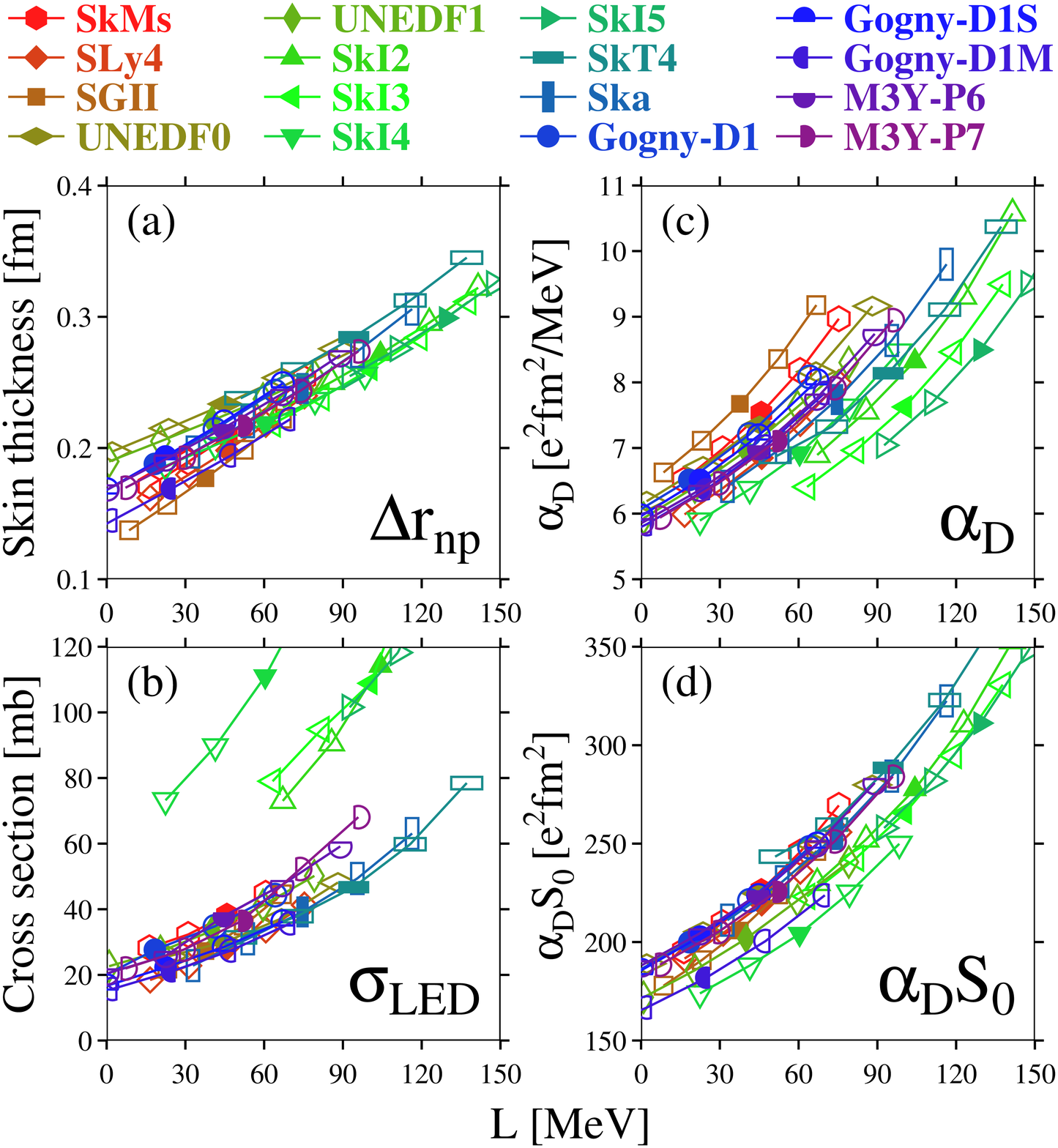}
\caption{(Color online) Correlations of the slope parameter $L$ with (a) the neutron skin thickness $\Delta r_{np}$, (b) the LED cross 
section $\sigma_\mathrm{LED}$, (c) the dipole polarizability $\alpha_D$ and (d) 
$\alpha_D S_0$ of $^{132}$Sn. See text for details.
}
\label{corr.132Sn}
\end{center}
\end{figure}

Figure~\ref{corr.132Sn} shows correlations of $L$ with the 
neutron skin thickness $\Delta r_{np}$, the LED cross section 
$\sigma_\mathrm{LED}$, the dipole polarizability $\alpha_D$ and 
$\alpha_D S_0$ in $^{132}$Sn, obtained by the HF+RPA calculations.
Effective interactions are distinguished by colors and symbols,
as listed in the upper part of the figure.
Results with $V_L=0$ are represented by full symbols,
while those with their $V_L \neq 0$ variants by open symbols.
The results of the same class of interactions
are connected by lines so as to show the CSIs.

It is seen in Fig.~\ref{corr.132Sn}(a) that $\Delta r_{np}$ correlates well with $L$.
Indeed, we obtain $R[L,\Delta r_{np}]=0.959$.
This correlation is well expressed by a linear function,
as $\Delta r_{np} = 0.00114 L + 0.160$\,fm
with standard deviation 0.014\,fm,
by assuming the unit of $L$ to be MeV.
With respect to the CSI, 
the three interactions UNEDF0, UNEDF1 and SkI4 give slopes less than $1.0 \times 10^{-3}$\,fm/MeV,
while slopes of the other Skyrme interactions are
steeper than $1.2 \times 10^{-3}$\,fm/MeV
and those of the Gogny and M3Y interactions fall in the narrow range
$(1.15 \pm 0.05) \times 10^{-3}$\,fm/MeV.
The maximum (minimum) slope is $1.46$ $(0.83) \times 10^{-3}$\,fm/MeV
of SGII (UNEDF1),
which deviates by 30\,\% from the value fitting all the interactions
(i.e. $1.14 \times 10^{-3}$\,fm/MeV).
We note that slopes of the CSI stay around $1.14 \times 10^{-3}$\,fm/MeV within 10\,\%
in more than half of the interactions.
The CDI (correlations among the interactions with $V_L=0$)
is strong as well, having $R[L,\Delta r_{np}]=0.939$.
Thus $L$ can be well constrained by $\Delta r_{np}$ in $^{132}$Sn
if it is measured precisely.
The standard deviation 0.014\,fm is converted to an uncertainty of 12\,MeV for $L$.

Correlations between $L$ and $\sigma_\mathrm{LED}$ are shown in Fig.~\ref{corr.132Sn}(b). 
We discard the $\sigma_\mathrm{LED}$ results
in the case that $S^{(E1)}(\omega)$ has two peaks in the LED region,
because we cannot unambiguously determine $\omega_\mathrm{dip}$
at which the LED and GDR regions are separated,
and $\omega_\mathrm{dip}$ may change discontinuously by changing $V_L$
even if we adopt a certain definition.
Four interactions SkI2, SkI3, SkI4 and SkI5 and their variants produce quite large $\sigma_\mathrm{LED}$,
which clearly deviate from the results of the other interactions.
The CSI are not similar even within these four classes of interactions.
It is noted that the Gogny and M3Y interactions yield correlations 
similar to the Skyrme interactions other than the above SkI series.
If we ignore the results of the SkI series,
$R[L,\sigma_\mathrm{LED}]=0.928$ is obtained
and $\sigma_\mathrm{LED}$ can be fitted to a linear function of $L$
as $\sigma_\mathrm{LED}= 0.399 L + 15.4$\,mb\,MeV with the standard deviation 5.0\,mb\,MeV.
When we fit $\sigma_\mathrm{LED}$ by a quadratic function,
we obtain $\sigma_\mathrm{LED}= 0.00138 L^2 + 0.238 L + 18.7$\,mb\,MeV with the standard deviation 4.7\,mb\,MeV.
Compared with Ref.~\cite{Carbone10} (see Fig.~2(b) of Ref.~\cite{Carbone10}),
the slope of the linear function is smaller by a factor $\sim 2$.
This discrepancy can be interpreted as follows:
In Ref.~\cite{Carbone10}, the CDI of $\sigma_\mathrm{LED}$ with $L$ has been investigated
via 19 Skyrme interactions and 7 relativistic effective Lagrangians which cover $L= 0 - 130$\,MeV.
Among them, seven relativistic Lagrangians and three Skyrme interactions
SkI2, SkI3 and SK255~\cite{Agrawal03},
all of which give $L \gtrsim 100$\,MeV,
seem to behave differently from the other interactions.
The high weight (10 out of the 26 interactions) of these large-$L$ interactions
leads to the steep slope in Ref.~\cite{Carbone10}.
If we exclude the results of SkI2, SkI3, SK255 and the RMF
in Fig.~2(b) of Ref.~\cite{Carbone10} and refit the others to a linear function,
the slope is compatible with our result.
However, with ambiguity in the definition of $\sigma_\mathrm{LED}$ and large deviation by certain interactions,
we conclude that $\sigma_\mathrm{LED}$ is currently unsuitable for constraining $L$.

Figure~\ref{corr.132Sn}(c) shows the $\alpha_D$-$L$ relations.
Despite the relatively large value of $R[L,\alpha_D]=0.90$,
the lines representing the CSI are widely scattered.
This indicates that the CDI behaves differently from the CSI.
If all the results of $\alpha_D$ are fitted to a linear function,
we obtain $\alpha_D = 0.0261 L + 5.94$\,e$^2$fm$^2$/MeV
with the standard deviation 0.53\,e$^2$fm$^2$/MeV.
However, the slopes given by the CSI are significantly larger;
$0.031 - 0.051$\,e$^2$fm$^2$/MeV$^2$ with the Skyrme interactions,
$\sim 0.027$ with the Gogny interactions
and $\sim 0.033$ with the M3Y interactions.
The intercepts are also distributed in as wide range as 2.14 -- 6.15\,e$^2$fm$^2$/MeV.
It is thus important to take into account the CSI and the CDI simultaneously.
The $\alpha_D$-$L$ correlation might look good when we pay attention only to the CSI like in the previous covariance analysis,
and likewise to the CDI.
However, there exists notable difference between the CSI and the CDI.
It is not necessarily suitable to constrain $L$ only by $\alpha_D$.

The $\alpha_DS_0$-$L$ correlations are shown in Fig.~\ref{corr.132Sn}(d).
The strong correlation between $\alpha_DS_0$ and $L$ is clearly seen.
The correlation coefficient is $R[L,\alpha_DS_0]=0.953$.
The linear fitting gives $\alpha_D S_0 = 1.13 L + 170$\,e$^2$fm$^2$ with the standard deviation 15\,e$^2$fm$^2$,
which corresponds to 13\,MeV uncertainty of $L$, 
and the quadratic fit gives $\alpha_D S_0 = 0.00393 L^2 + 0.617 L + 180$\,e$^2$fm$^2$
with the standard deviation 13\,e$^2$fm$^2$.
Even if we restrict ourselves to the CDI by setting $V_L=0$,
the correlations have similar behavior;
$R[L,\alpha_DS_0]=0.947$,
the fitted linear function is $\alpha_D S_0 = 1.08 L + 169$\,e$^2$fm$^2$
with the standard deviation 11\,e$^2$fm$^2$,
and the quadratic function is $\alpha_D S_0 = 0.00208 L^2 + 0.773 L + 177$\,e$^2$fm$^2$
with the standard deviation 11\,e$^2$fm$^2$.
As pointed out in Ref.~\cite{Lattimer13},
$S_0$ has positive correlation with $L$ among the interactions with $V_L=0$.
This helps $\alpha_DS_0$ to correlate with $L$ better than $\alpha_D$ alone.
Thus $\alpha_D S_0$ will be useful in constraining $L$,
although it requires precise assessment of $S_0$.

We have investigated the correlations of $\Delta r_{np}$, $\sigma_\mathrm{LED}$, 
$\alpha_D$ and $\alpha_D S_0$ in $^{132}$Sn with $L$,
and have found that $\Delta r_{np}$ and $\alpha_DS_0$ are promising
for constraining $L$.

\subsection{Nucleus-dependence}

The correlations between $L$ and observables related to the neutron skin
were discussed mainly in $^{68}$Ni, $^{132}$Sn and $^{208}$Pb in the previous studies. 
We next consider nucleus-dependence of the $\Delta r_{np}$-$L$ and $\alpha_D S_0$-$L$ correlations. 

\begin{figure}[tb]
\begin{center}
\includegraphics[width=0.4990\textwidth,keepaspectratio]{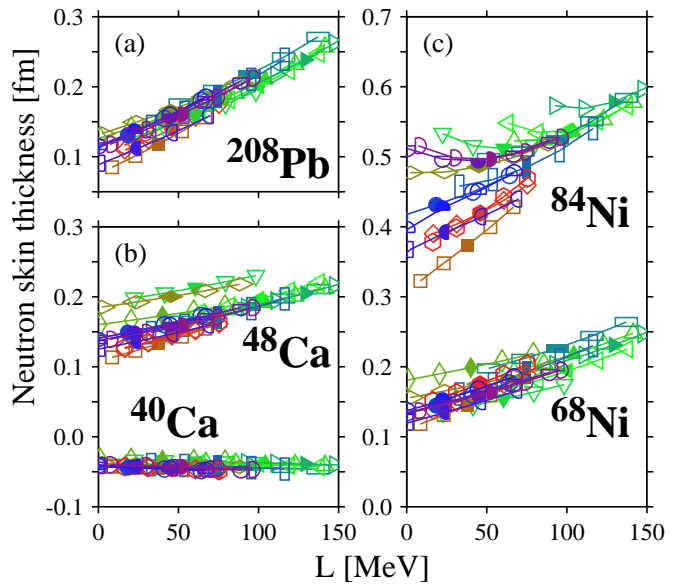}
\caption{(Color online) Correlations of $\Delta r_{np}$ and $L$
in (a) $^{208}$Pb, (b) $^{40,48}$Ca and (c) $^{68,84}$Ni. 
See Fig.~\ref{corr.132Sn} for colors and symbols.
}
\label{L.skin}
\end{center}
\end{figure}

We have calculated the $\Delta r_{np}$-$L$ correlations in doubly-magic nuclei
and in nearly-doubly-magic nuclei,
$^{16,22,24}$O, $^{40,48,54,70}$Ca, $^{68,78,84}$Ni, $^{132,140,176}$Sn and $^{208}$Pb,
some of which are plotted in Fig.~\ref{L.skin}.
In $^{208}$Pb, $\Delta r_{np}$ correlates well to $L$,
giving $R[L,\Delta r_{np}(\mbox{$^{208}$Pb})]=0.965$.
This result is consistent with that reported in Ref.~\cite{RocaMaza11}.
The linear function obtained by fitting 
is $\Delta r_{np}(\mbox{$^{208}$Pb}) = 0.00107 L + 0.103$\,fm 
with the standard deviation 0.013\,fm, being equivalent to 12\,MeV uncertainty of $L$.
The slope of the fitted function is smaller by $\sim 30 \%$ than that of Ref.~\cite{RocaMaza11}.
This discrepancy is again attributed to contribution of the RMF results with $L \gesim 100$\,MeV,
because they increase the slope in Fig.~3 of Ref.~\cite{RocaMaza11}.
Still the $\Delta r_{np}$-$L$ correlation in $^{208}$Pb is so strong
to be promising for getting constraint on $L$.
The $\Delta r_{np}$-$L$ correlation gradually becomes the weaker
for the lighter nuclei.
Notice that the steeper slope in the linear function
tends to make the correlation coefficient the larger.
When errors in experimental data are taken into consideration,
a steep slope is further advantageous in constraining $L$.
As mentioned above, we obtain $R[L,\Delta r_{np}(\mbox{$^{132}$Sn})]=0.959$. 
In $^{68}$Ni, $R[L,\Delta r_{np}(\mbox{$^{68}$Ni})]=0.901$
and the linear fitting gives $\Delta r_{np}(\mbox{$^{68}$Ni}) = 0.000761 L + 0.133$\,fm.
In $^{48}$Ca, the correlation coefficient drops
to $R[L,\Delta r_{np}(\mbox{$^{48}$Ca})]=0.785$
and the linear fitting results in $\Delta r_{np}(\mbox{$^{48}$Ca}) = 0.000546 L + 0.138$\,fm.
In the $Z=N$ nuclei $^{16}$O and $^{40}$Ca,
the calculated $\Delta r_{np}$'s are almost independent of $L$.
The $\Delta r_{np}$-$L$ correlation also becomes weak in drip-line nuclei
such as $^{84}$Ni, as shown in Fig.~\ref{L.skin}(c).
$\Delta r_{np}$ is strongly affected by the spatial extension of the 
loosely-bound neutron orbits around the neutron Fermi level.
In nuclei near the neutron drip line,
the additional term introduced in Eq.~(\ref{VL}) with negative $V_L$,
which lowers $L$, lifts up the neutron Fermi level
and makes the loosely-bound orbits extend significantly.
This effect is connected to the neutron halo
which may irregularly increase $\Delta r_{np}$. 
This mechanism makes the $\Delta r_{np}$-$L$ correlation weaker
in neutron drip-line nuclei, as is seen in $^{22,24}$O, $^{70}$Ca and $^{176}$Sn.

Therefore, the $\Delta r_{np}$ in heavy nuclei distant from the drip line
may be appropriate in constraining $L$.
Measurement on $^{208}$Pb seems to provide one of the best possibilities in this respect.
However, despite great efforts and progress,
it is not yet easy to experimentally determine $\Delta r_{np}(\mbox{$^{208}$Pb})$
with good precision.
It should also be kept in mind that
the $\Delta r_{np}$-$L$ correlation has been investigated only phenomenologically.
Without support from quantitatively reliable theories,
cross checks from other nuclei and/or other quantities are important.


\begin{table*}
\caption{Correlation coefficient $R[L,\alpha_DS_0]$,
and the optimized values of the coefficients $a$ and $b$
when the calculated results are fitted as $\alpha_DS_0=aL+b$,
with the standard deviation $\sigma$ of the fitting.} 
\begin{ruledtabular}
\begin{tabular}{c|cccc}
nucleus ($^AZ$)  & $R[L,\alpha_DS_0]$  & $a$ [e$^2$fm$^2$/MeV] & $b$ [e$^2$fm$^2$] & $\sigma$ [e$^2$fm$^2$] \\ \hline
 $^{16}$O   &    0.848  &    0.068  &    9.6  &    1.8   \\
 $^{24}$O   &    0.628  &    0.058  &    9.8  &    1.5   \\ \hline
 $^{40}$Ca  &    0.881  &    0.210  &   33.4  &    4.7   \\
 $^{48}$Ca  &    0.898  &    0.242  &   39.6  &    5.0   \\
 $^{54}$Ca  &    0.857  &    0.331  &   69.1  &    8.4   \\ \hline
 $^{68}$Ni  &    0.929  &    0.448  &   70.5  &    7.6   \\
 $^{84}$Ni  &    0.662  &    0.574  &  142.7  &   27.7   \\ \hline
 $^{132}$Sn &    0.953  &    1.130  &  170.1  &   15.2   \\
 $^{140}$Sn &    0.934  &    1.354  &  213.6  &   21.9   \\ \hline
 $^{208}$Pb &    0.927  &    1.864  &  336.8  &   31.8   \\
\end{tabular}
\end{ruledtabular}
\label{tab:pol.L}
\end{table*}

\begin{figure}[tb]
\begin{center}
\includegraphics[width=0.4990\textwidth,keepaspectratio]{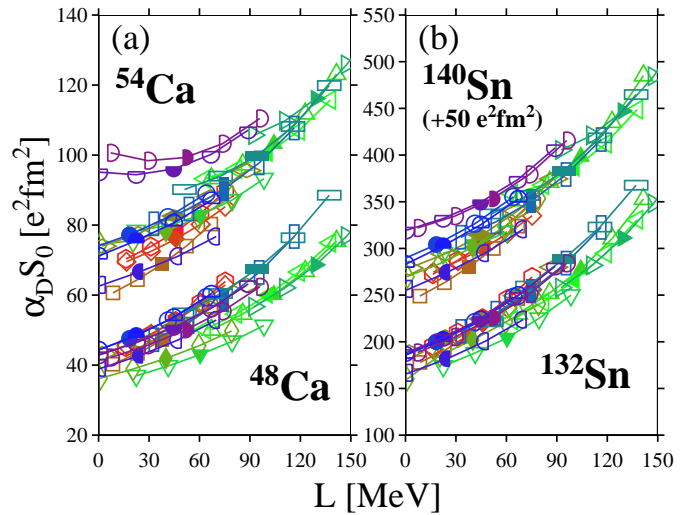}
\caption{(Color online) $\alpha_DS_0$-$L$ correlations in (a) $^{48,54}$Ca and 
(b) $^{132,140}$Sn. Data of $^{140}$Sn is shifted by 50 e$^2\mbox{fm}^2$ 
to accommodate them with the results of $^{132}$Sn in a single plot.
See Fig.~\ref{corr.132Sn} for colors and symbols.
}
\label{Ca.Sn}
\end{center}
\end{figure}

Let us turn to nucleus-dependence of the $\alpha_D S_0$-$L$ correlation.
Because of the energy denominator in Eq.~(\ref{polarizability}),
$\alpha_D S_0$ is rather sensitive to the LED,
which emerges and grows up beyond the magic numbers $N=14, 28, 50$
and $82$~\cite{Inakura11,Ebata14}.
We expect that $\alpha_D S_0$ correlates better with $L$
as the LED develops in the neutron-rich nuclei.
In Table~\ref{tab:pol.L} we list $R[L,\alpha_DS_0]$
for the stable doubly-magic nuclei and neutron-rich nuclei having well-developed LED, 
$^{16,24}$O, $^{40,48,54}$Ca, $^{68,84}$Ni, $^{132,140}$Sn and $^{208}$Pb.
The optimized values of the coefficients $a$ and $b$
when the calculated results are fitted as $\alpha_DS_0=aL+b$,
with the standard deviation $\sigma$ of the fitting,
are shown as well.

The left panel of Fig.~\ref{Ca.Sn} illustrates
how the LED affects the $\alpha_DS_0$-$L$ correlation,
by comparing the results of $^{54}$Ca with those of $^{48}$Ca.
From $^{48}$Ca to $^{54}$Ca, $\alpha_DS_0$ becomes larger
and the slope of the fitted linear function becomes steeper (0.242 to 0.331\,e$^2$fm$^2$/MeV).
The steep slope is expedient for constraining $L$ from experiment.
The LED emergence and development contributes to the $\alpha_DS_0$-$L$ correlation.
However, in $^{54}$Ca the $\alpha_D S_0$-$L$ relation of the M3Y-P6 and P7 interactions 
deviates significantly from that of the other interactions,
while such deviation is not found in $^{48}$Ca.
As a result, we obtain $R[L,\alpha_DS_0(\mbox{$^{54}$Ca})]=0.86$,
smaller than $R[L,\alpha_DS_0(\mbox{$^{48}$Ca})]=0.90$.
This is mainly because the M3Y-P6 and P7 interactions produce
higher neutron Fermi level than the other interactions in $^{54}$Ca,
and generate the neutron halo when we take $V_L<0$.
As in $\Delta r_{np}$, presence of the halo disturbs the correlation,
since the halo may produce large LED and thereby causes large $\alpha_D$.
It can be confirmed experimentally whether or not $^{54}$Ca is a halo nucleus.
Suppose that the neutron halo is ruled out,
$^{54}$Ca can be a candidate to constrain $L$ from $\alpha_DS_0$.
Excluding the M3Y interactions,
we obtain $R[L,\alpha_DS_0(\mbox{$^{54}$Ca})]=0.96$
and steeper slope (0.40\,e$^2$fm$^2$/MeV) in the linear fitting.
Also for $^{68,84}$Ni,
whereas the slope obtained by the linear fitting becomes steeper in $^{84}$Ni,
$R[L,\alpha_DS_0(\mbox{$^{84}$Ni})]=0.66$ is small
because of the neutron halo.
Similar trend is seen in the drip-line nucleus $^{24}$O.
Before applying $\alpha_DS_0$ in a certain nucleus for constraining $L$,
it should be confirmed that the nucleus does not have halo.

The correlation coefficients are high both in $^{132,140}$Sn,
$R[L,\alpha_DS_0(\mbox{$^{132}$Sn})]=0.95$
and $R[L,\alpha_DS_0(\mbox{$^{140}$Sn})]=0.93$,
as presented in the right panel of Fig.~\ref{Ca.Sn}.
Although $R[L,\alpha_DS_0]$ slightly decreases from $^{132}$Sn to $^{140}$Sn,
the slope becomes steeper.
Both nuclei are suitable for constraining $L$ from $\alpha_DS_0$,
if $\alpha_D$ is accessible in future experiments.

The $\alpha_D S_0$-$L$ correlation in $^{208}$Pb has been calculated in Ref.~\cite{RocaMaza13} 
employing Skyrme interactions and relativistic Lagrangians, 
and the linear fitting gives the slope $a = 2.3$\,e$^2$fm$^2$/MeV and the intercept $b = 333$\,e$^2$fm$^2$.
Compared with our result, the intercept is almost equal but the slope is steeper.
Another result of the $\alpha_D S_0$-$L$ relation is available from Ref.~\cite{Vretenar12}, 
in which only $\alpha_D$-$L$ correlation is calculated with a family of relativistic Lagrangian.
We can see the $\alpha_D S_0$-$L$ correlation using those results.
The fitted linear function representing the $\alpha_D S_0$-$L$ correlation of Ref.~\cite{Vretenar12} has the 
slope $a \sim 2.9$\,e$^2$fm$^2$/MeV and the intercept $b \sim 310$\,e$^2$fm$^2$. 
The slope is again steeper than our result while the intercept is compatible.
Therefore, the currently available RMF results increase the slope but have small impact on the intercept of 
the $\alpha_D S_0$-$L$ relation.


\subsection{Comparison with droplet model estimation}

The $\alpha_D S_0$-$L$ correlation has been suggested in Ref.~\cite{RocaMaza13}
based on the droplet model under some assumptions.
The relation of $\alpha_D S_0$ and $L$ reads
\begin{equation}
(\alpha_D S_0)_\mathrm{DM} \sim \frac{\pi e^2}{54}A \langle r^2 \rangle \left[ 1 + \frac{5}{9} \frac{L}{S_0} \frac{\rho_0 - \rho_A}{\rho_0} \right] \,,
\label{aDS0DM}
\end{equation}
where $\rho_A \sim 0.1$\,fm$^{-3}$~\cite{Centelles09,Wang15}.
While we have phenomenologically confirmed the $\alpha_D S_0$-$L$ correlation
in preceding sections,
it deserves investigating validity of this relation. 
For a given interaction and a nucleus
we evaluate $(\alpha_D S_0)_\mathrm{DM}$ from Eq.~(\ref{aDS0DM}), 
and compare them to $\alpha_DS_0$ obtained from the HF+RPA calculations. 

\begin{table*}
\caption{Saturation density $\rho_0$, incompressibility of symmetry 
nuclear matter $K_\infty$, symmetry energy $S_0$, slope parameter $L$, 
incompressibility of symmetry energy $K_\mathrm{sym}$, surface stiffness 
parameter $Q$ and $x_A$ multiplied by the mass dependence,
given by the Skyrme, Gogny and M3Y interactions with $V_L=0$.}
\begin{ruledtabular}
\begin{tabular}{c|c|c|c|c|c|c|c}
     & $\rho_0$ [$\mathrm{fm}^{-3}$] & $K_\infty$ [MeV] & $S_0$ [MeV] & $L$ [MeV] & $K_\mathrm{sym}$ [MeV] & $Q$ [MeV] & $x_A \times A^{1/3}$ \\ \hline
 SkM$^\ast$  &  0.160  &   216.4 &  30.0  &  45.8 & -155.8  &  35.9  &   1.88 \\
 SLy4        &  0.160  &   229.9 &  32.0  &  45.9 & -119.7  &  42.8  &   1.68 \\
 SGII        &  0.158  &   214.5 &  26.8  &  37.7 & -145.8  &  33.9  &   1.78 \\
 UNEDF0      &  0.160  &   229.8 &  30.5  &  45.1 & -189.6  &  35.8  &   1.92 \\
 UNEDF1      &  0.159  &   219.8 &  29.0  &  40.0 & -179.4  &  35.8  &   1.82 \\
 SkI2        &  0.157  &   240.7 &  33.4  & 104.3 &   70.6  &  26.6  &   2.82 \\
 SkI3        &  0.158  &   258.0 &  34.8  & 100.5 &   72.9  &  30.2  &   2.60 \\
 SkI4        &  0.160  &   247.7 &  29.5  &  60.4 &  -40.6  &  31.9  &   2.08 \\
 SkI5        &  0.156  &   255.6 &  36.6  & 129.3 &  159.4  &  27.3  &   3.02 \\
 SkT4        &  0.159  &   262.9 &  35.5  &  94.1 &  -24.5  &  30.6  &   2.60 \\
 Ska         &  0.155  &   235.3 &  32.9  &  74.6 &  -78.4  &  31.5  &   2.35 \\ \hline
 D1          &  0.166  &   229.4 &  30.7  &  18.4 & -274.6  &  41.4  &   1.67 \\
 D1S         &  0.163  &   202.9 &  31.1  &  22.4 & -241.5  &  41.0  &   1.71 \\
 D1M         &  0.165  &   225.0 &  28.6  &  24.8 & -133.2  &  41.0  &   1.57 \\ \hline
 M3Y-P6      &  0.163  &   239.7 &  32.1  &  44.6 & -165.3  &  31.8  &   2.27 \\
 M3Y-P7      &  0.163  &   254.7 &  31.7  &  51.5 & -127.8  &  29.2  &   2.45 \\
\end{tabular}
\end{ruledtabular}
\label{tab:interaction}
\end{table*}

One of the assumptions in the droplet model is~\cite{Myers69,Brack85} 
\begin{equation}
x_A \equiv \frac{9}{4}\frac{S_0}{Q}A^{-1/3} \ll 1 \,,
\end{equation}
where $Q$ is the surface stiffness coefficient
connected with the nuclear surface symmetry energy~\cite{Danielewicz09}.
For the droplet model estimation (Eq.~(\ref{aDS0DM})) to be justified,
$x_A$ should be sufficiently small.
In evaluating $x_A$,
we use an approximate expression for $Q$~\cite{Chen11},
\begin{equation}
Q \sim \frac{9 S^2_0}{8a} \left( \frac{4}{3}\pi \rho_0 \right)^{-1/3} \left( L - \frac{K_\mathrm{sym}}{12}\right)^{-1}\,,
\end{equation}
instead of calculating $Q$ in the asymmetric semi-infinite 
nuclear matter~\cite{Myers69,Brack85,Danielewicz09,Centelles98}.
Here $a$\,($\sim 0.55$\,fm) is the diffuseness of the symmetric semi-infinite nuclear matter~\cite{Danielewicz09}
and $K_\mathrm{sym}$ is the 2nd derivative of the symmetry energy with respect to the density at the saturation point.
The calculated $x_A$ values are listed in Table~\ref{tab:interaction},
accompanying $\rho_0$, $K_\infty$, $S_0$, $L$, $K_\mathrm{sym}$ and $Q$.

\begin{figure}[tb]
\begin{center}
\includegraphics[width=0.4990\textwidth,keepaspectratio]{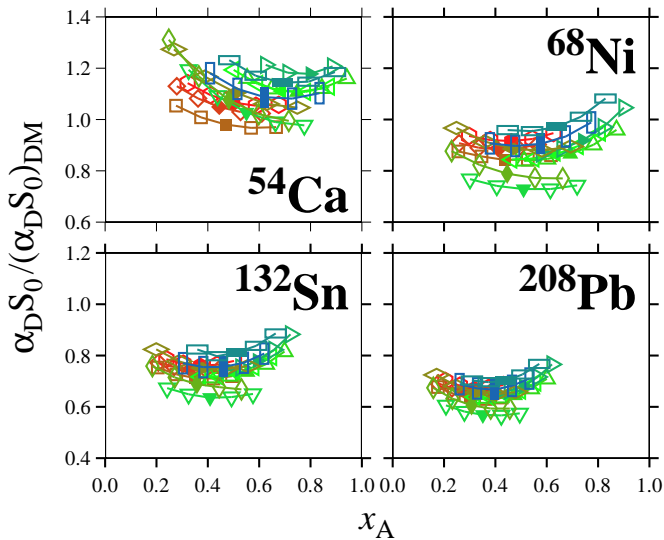}
\caption{(Color online) Ratio of $\alpha_DS_0$
obtained from the HF+RPA calculations to the droplet model estimate 
in $^{54}$Ca, $^{68}$Ni, $^{132}$Sn and $^{208}$Pb.
See Fig.~\ref{corr.132Sn} for colors and symbols.
}
\label{aDS0Ratio}
\end{center}
\end{figure}

Figure~\ref{aDS0Ratio} shows the ratio of $\alpha_D S_0$ calculated with HF+RPA 
to $(\alpha_D S_0)_\mathrm{DM}$ in $^{54}$Ca, $^{68}$Ni, $^{132}$Sn and $^{208}$Pb.
The ratios clearly deviate from unity even for small $x_A$.
Although we have phenomenologically confirmed the $\alpha_D S_0$-$L$ correlation,
the droplet model is not necessarily appropriate for justifying the correlation.

\section{Conclusion}
\label{sec.conclusion}

We have investigated the correlations of $L$ with the following four quantities;
the neutron skin thickness $\Delta r_{np}$,
the cross section of the low-energy dipole (LED) mode $\sigma_\mathrm{LED}$,
the dipole polarizability $\alpha_D$,
and the product of $\alpha_D$ and the symmetry energy $S_0$.
In order to directly compare them 
and to unravel disorder in observables constraining $L$,
we have simultaneously discussed the correlations derived from different interactions 
(CDI) and the correlation in a single class of interactions (CSI).
For the latter we introduce an additional term to each interaction,
which enables us to control the value of $L$
without influencing SNM EoS and $S_0$.

The $\Delta r_{np}$ correlates almost linearly with $L$ in heavy nuclei,
although there remains slight interaction-dependence
as recognized via comparison with the results in Ref.~\cite{RocaMaza11}.
The $\sigma_\mathrm{LED}$-$L$ correlation has a significant interaction-dependence.
Together with ambiguity in its definition,
$\sigma_\mathrm{LED}$ is not recommended to constraining $L$.
In the $\alpha_D$-$L$ correlation,
we have found that the CSI and the CDI behave differently.
It is not reasonable to constrain $L$ only from $\alpha_D$.
The $\alpha_DS_0$-$L$ correlation works well for narrowing down $L$.
The $\Delta r_{np}$ and $\alpha_DS_0$ are promising for constraining $L$, 
though with $\sim 12$\,MeV uncertainty.

The nucleus-dependence of the $\Delta r_{np}$-$L$ and $\alpha_D S_0$-$L$ 
correlations has also been discussed.
While the neutron halo makes the correlations weak,
these correlations are strong in neutron-rich medium- or heavy-mass nuclei without neutron halo.
Except neutron-halo nuclei,
the LED makes the $\alpha_D S_0$-$L$ correlation strong
and the slope of the linear function steep,
to which the HF+RPA results are well fitted.
Consequently, the neutron-rich nuclei having well-developed LED
(e.g. $^{54}$Ca and $^{140}$Sn) are good candidates for obtaining 
the constraint on $L$,
as well as the doubly magic nuclei $^{132}$Sn and $^{208}$Pb.

\section*{Acknowledgments}

We thank K. Iida for fruitful discussions.
This work is financially supported as Grant-in-Aid for Scientific 
Research on Innovative Areas, No. 24105008, by The Ministry of Education, 
Culture, Sports, Science and Technology, Japan.
A part of numerical calculations 
were performed on HITAC SR16000s at IMIT in Chiba University, ITC in 
University of Tokyo, IIC in Hokkaido University, and YITP in Kyoto 
University.

\end{document}